\documentclass[floatfix,twocolumn,aps,prl,showpacs,amsmath,amssymb]{revtex4}

\usepackage{epsfig}
\usepackage{color}

\begin{document}
\title{Electromagnetic wave scattering by Schwarzschild black holes}

\author{Lu\'is C. B. Crispino}
\email{crispino@ufpa.br}
\affiliation{Faculdade de F\'\i sica, Universidade Federal do
Par\'a, 66075-110, Bel\'em, PA,  Brazil}

\author{Sam R. Dolan}
\email{sam.dolan@ucd.ie}
\affiliation{School of Mathematical Sciences, University College
Dublin, Belfield, Dublin 4, Ireland}

\author{Ednilton S. Oliveira}
\email{ednilton@fma.if.usp.br}
\affiliation{Instituto de F\'\i sica, Universidade de S\~ao Paulo, 
CP 66318, 05315-970, S\~ao Paulo, SP, Brazil}

\date{\today}
\begin{abstract}

We analyze the scattering of a planar monochromatic electromagnetic wave 
incident upon a Schwarzschild black hole. We obtain accurate numerical 
results from the partial wave method for the electromagnetic scattering 
cross section, and show that they are in excellent agreement with analytical 
approximations. The scattering of electromagnetic waves is compared with 
the scattering of scalar, spinor and gravitational waves. We present a 
unified picture of the scattering of all massless fields for the first time.
\end{abstract}
\pacs{04.40.-b, 04.70.-s, 11.80.-m}

\maketitle

Black holes are thought to be efficient catalysts for the liberation of rest-mass energy. As such, black holes are implicated in the most energetic phenomena in the known universe (e.g. gamma ray bursts). On the other hand, after a turbulent youth, many black holes settle into a quiescent old age. Some estimates suggest there may be up to a billion quiescent stellar-mass black holes within our galaxy \cite{Begelman2003}. Their existence may be inferred from, for example, the transient lensing of background sources; a handful of events have so far been observed \cite{Bennett2002}. A possibility for  future consideration is that quiescent black holes may be indirectly identified from the `fingerprint' they leave on radiation that impinges upon them.

Over the last four decades, some clues about the properties of any such `fingerprint' have been uncovered. For example, a time-dependent perturbation incident upon a black hole will excite characteristic \emph{damped ringing} in response. The frequencies and decay rates of the ringing are linked to the well-studied quasinormal mode spectrum \cite{Kokkotas&Schmidt}. Black holes illuminated by long-lasting planar radiation will create interference patterns, and rotating black holes will create distinctive polarization patterns \cite{Dolan2006}. Both effects depend strongly on the ratio of horizon size to wavelength. Hence, it is conceivable that future gravitational-wave detectors may be able to identify the fingerprint from rapid and distinctive variations across a narrow frequency band. Nevertheless, inferring the presence of quiescent black holes from such clues must remain a challenge for future decades. 

Scattering by black holes is of foundational interest 
in both black hole physics~\cite{Futterman} and scattering theory~\cite{Gottfried&Yan2004}. 
Many authors have studied the simplest time-independent scenario, in which a black hole is subject to a long-lasting, monochromatic beam of radiation. Here, the key dimensionless quantity is the ratio 
$r_h /  \lambda$
where $r_{h}$ is the horizon size of the black hole, and $\lambda$ is the wavelength of the incident wave. 
The interference pattern depends also on the spin $s$ of the perturbing field, with $s = 0$, $1/2$, $1$ and $2$ corresponding to scalar, neutrino, electromagnetic and gravitational fields, respectively. 

To the best of our knowledge, the first paper 
outlining a calculation of wave scattering cross section 
in the spacetime of a black hole was published by 
Matzner~\cite{Matzner} in the late sixties.
Since then, planar wave scattering from black holes has received much attention, especially in Schwarzschild and Kerr spacetimes (see 
Refs.~\cite{Futterman, Andersson&Jensen, Frolov&Novikov} 
for comprehensive accounts on the subject). 
Let us briefly review a sample of the literature for the simplest case, the Schwarzschild black hole, for which the scattering of monochromatic fields of all spins ($s=0$, $1/2$, $1$ and $2$) has been studied through the years.
The case of scalar waves ($s=0$) was extensively studied by 
Sanchez~\cite{Sanchez1976, SanchezPRD}, 
both analytically and numerically, 
and an accurate numerical study was later 
performed by Andersson~\cite{Andersson1995}.
Fermion ($s=1/2$) scattering by a Schwarzschild black hole was 
the subject of a recent study~\cite{Dolan2006}, in which the authors also elucidated the effect of non-zero field mass. 
The case of electromagnetic waves ($s=1$) was studied analytically 
by Mashoon~\cite{Mashhoon1973} and Fabbri~\cite{Fabbri1975}, 
and some results were obtained in the low- and high-frequency
limits.
Gravitational waves ($s=2$) were the first to be studied in black hole 
scattering~\cite{Hildreth1964}, and are the subject of old \cite{Handler1980, Futterman} and new \cite{Dolan2008a, Dolan2008} works.

In this letter we present the first detailed numerical investigation of the scattering of an electromagnetic
plane wave by a Schwarzschild black hole. This work fills a gap in the literature, and complements  recent numerical studies of the scalar \cite{Glampedakis2001}, fermionic \cite{Dolan2006}, and gravitational \cite{Dolan2008} cases. We take this opportunity to present a unified picture of all four fields, for the first time. 

We use natural units with $c = G = 1$ and the metric
signature $(+ - - -)$.

%
%
The line element of Schwarzschild spacetime can be written as
\begin{equation}
 ds^2=f(r)dt^2-[f(r)]^{-1}dr^2-r^2(d\theta^2+\sin^2\theta d\phi^2),
 \label{Schw}
\end{equation}
where $f(r)= 1 - 2M/r$, with $M$ being the black hole mass. The
Schwarzschild solution describes static and chargeless black holes, 
with event horizon at $r_{h}=2M$.

The Lagrangian density of the electromagnetic field in the 
modified Feynman gauge is~\cite{CHM3PRD}
$$
\mathcal{L}
=-\sqrt{-g}\left[  \dfrac{1}{4}F_{\mu\nu}F^{\mu\nu}+\dfrac{1}{2}G^{2}\right]
$$
with $g=\det\left(  g_{\mu\nu}\right)$, 
$G\equiv\nabla^{\mu}A_{\mu}+K^{\mu}A_{\mu}~$ and 
$
K^{\mu}=\left(  0,df/dr,0,0\right)  \text{.}%
$
The equations of motion are found to be
\begin{equation}
\nabla_{\nu} F^{\mu \nu} + \nabla^{\mu} G - K^{\mu} G = 0.
\label{ME}
\end{equation}

The two physical polarizations in Schwarzschild spacetime can be written as
\begin{eqnarray}
& & A_{\mu}^{ (I\omega lm) }=(0\,,\,
\dfrac{{\varphi}_{\omega l}^{I}\left(  r\right)}{r^2}  ~Y_{lm},\dfrac
{f}{l\left(  l+1\right)  }\dfrac{d}{dr}\left[  \varphi_{\omega l}^{I}\left(
r\right)  \right]   
\nonumber
\\
& & \times \partial_{\theta}Y_{lm},
\dfrac{f}{l\left(  l+1\right)  }\dfrac{d}{dr}\left[  \varphi_{\omega l}%
^{I}\left(  r\right)  \right]  \partial_{\phi}Y_{lm})e^{-i\omega t}\text{,}%
\label{modo I}
\end{eqnarray}
\begin{equation}
A_{\mu}^{ (II\omega lm) }=\left(  0,0,\varphi_{\omega l}^{II}\left(  r\right)  Y_{\theta
}^{lm},\varphi_{\omega l}^{II}\left(  r\right)  Y_{\phi}^{lm}\right)  e^{-i\omega
t}\text{,} \label{modo II}%
\end{equation}
with $\omega>0$, and $l\geqslant 1$.
(See, e. g., Ref.~\cite{CHM3PRD} for a discussion on the possible 
solutions of Eq.~(\ref{ME}).)
In Eqs.~(\ref{modo I}) and~(\ref{modo II}), $Y_{lm}$ and $Y_{a}^{lm}$
($a=\theta, \phi$)
are the scalar and vector spherical harmonics~ \cite{AHCQG},
respectively, and ${\varphi}_{\omega l}^{\lambda}\left(  r\right)$
satisfy the following equation
\begin{equation}
 \left(  \omega^{2}-V(r)\right)  \varphi_{\omega l}^{\lambda 
}\left(  r\right)
 +f\dfrac{d}{dr}\left(  f\dfrac{d}{dr}\varphi_{\omega
 l}^{\lambda}\left(
 r\right)\  \right)  =0 \text{,}
 \label{Radial eq}%
\end{equation}
with $\lambda=I,II$, where the effective potential is $V(r)=f[l(l+1)/r^2]$.

To evaluate the solutions of the Eq.~(\ref{Radial eq}) in 
the asymptotical limits 
we use the Wheeler coordinate, defined as
$ x=\allowbreak r+r_{s}\ln\left(r/r_{s}-1\right)$,
and rewrite Eq.~(\ref{Radial eq}) as
\begin{equation}
\left(  \omega^{2}-V\right)  \varphi_{\omega l}^{\lambda}\left( 
x\right)
+\dfrac{d^{2}}{dx^{2}}\left(  \varphi_{\omega l}^{\lambda}\left(
x\right)
\right)  =0\text{.}
\label{Radial eq in x}%
\end{equation}

For the computation of the scattering cross section 
we need only to consider modes incoming from 
the past null infinity ${\cal J}^{-}$.
For these modes, the asymptotic solutions of 
Eq.~(\ref{Radial eq in x}) are~\cite{CHMO}
\begin{equation}
 \varphi_{\omega l}^{\lambda} (x) \approx A_{\omega l}^{\lambda}
T_{\omega l}^{\lambda} e^{-i\omega x},
 \label{varphi at horizon}
\end{equation}
for $x\rightarrow - \infty$ ($r\rightarrow r_s$) and
\begin{equation}
 \frac{\varphi_{\omega l}^{\lambda}(x)}{\omega x} \approx A_{\omega
l}^{\lambda} \left[
(-i)^{l+1} h_{l}^{(1)*}(\omega x) + R_{\omega l}^{\lambda}
i^{l+1} h_{l}^{(1)}(\omega x) \right],
\label{varphi at infinity}
\end{equation}
for $x \gg r_{s}$ ($r\gg r_{s}$).
Here
$h_{l}^{\left(  1\right)  }(x)$
denote the spherical Bessel functions of the third kind \cite{Abramo},
and $|R_{\omega l}^{\lambda}|^2$ and $|T_{\omega l}^{\lambda}|^2$
are the reflexion and transmission coefficients, respectively, which satisfy
$|R_{\omega l}^{\lambda}|^{2}+|T_{\omega l}^{\lambda}|^{2} = 1$.
$A_{\omega l}^{\lambda}$ is a normalization constant which is not
important for the scattering properties.

The phase shifts are related to the reflexion coefficient by
\begin{equation}
 e^{2i\delta_{l}^{\lambda}(\omega)} = (-1)^{l+1}R_{\omega l}^{\lambda}\,.
\label{Phase shift}
\end{equation}
For Schwarzschild black holes, 
the phase shifts of the two different physical polarizations are
the same, i. e. $\delta_{l}^{I}(\omega) = \delta_{l}^{II}(\omega) = 
\delta_{l}(\omega)$~\cite{Fabbri1975}.

The differential electromagnetic
scattering cross section is~\cite{Fabbri1975sign} 
\begin{eqnarray}
 \frac{d\sigma}{d\Omega} &=& \frac{1}{4\omega^2} \left|\sum
 \limits_{l=1}^{\infty}\frac{2l+1}{l(l+1)} e^{+2i\delta_{l}(\omega)}
 \left[\frac{P_{l}^{1}(\cos\theta)}{\sin\theta} \right. \right.
 \nonumber
 \\
 &+& \left. \left. \frac{d}{d\theta}P_{l}^{1}(\cos\theta) \right] \right|^2\,,
 \label{electro_scs}
\end{eqnarray}
where $P_{l}^{m}(\cos\theta)$ are the associated Legendre functions.
Note that Eq.~(\ref{electro_scs}) takes into account the contributions 
from the two physical polarizations, and it 
is valid for both linearly and circularly polarized waves.
The polarization properties of the initial wave remain unchanged 
in the scattering by non-rotating black holes~\cite{Mashhoon1973}.

%
%

\begin{figure}
\includegraphics[scale=1]{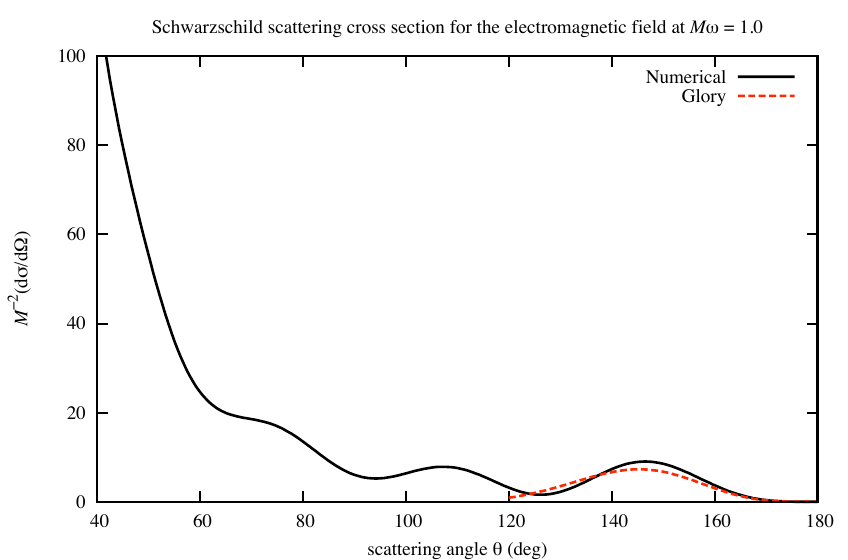}
\includegraphics[scale=1]{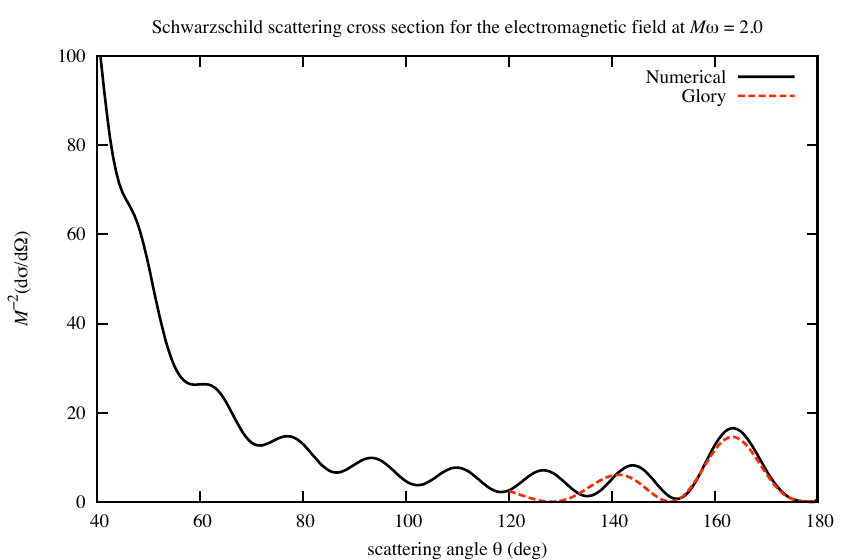}
\includegraphics[scale=1]{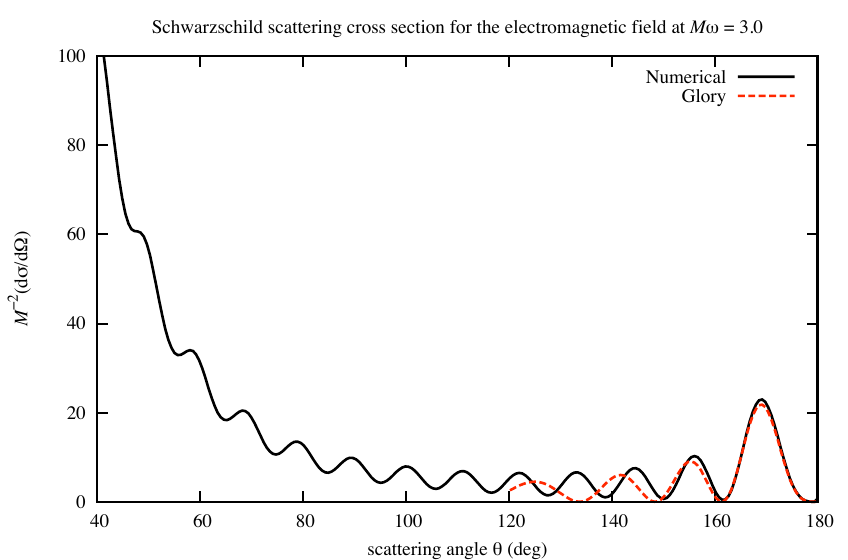}
\includegraphics[scale=1]{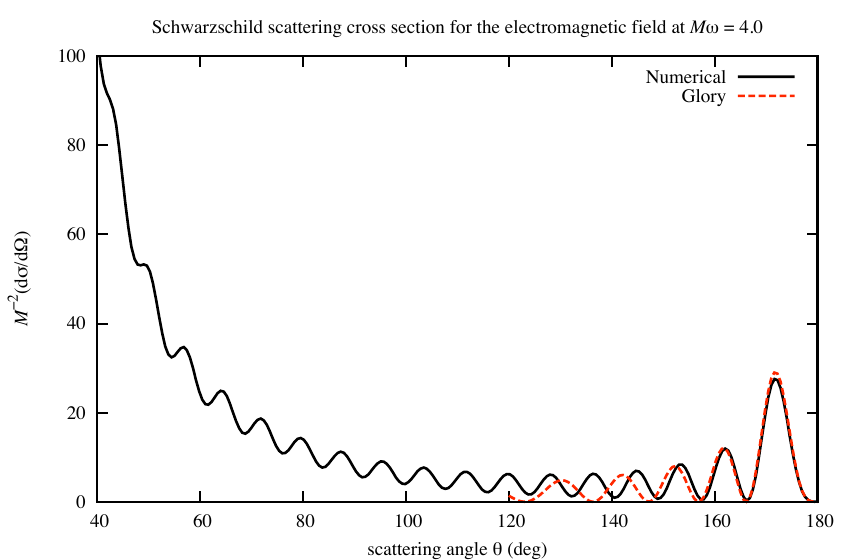}
\caption{Electromagnetic scattering cross section of
Schwarzschild black holes for different choices of $M \omega$. 
We compare our numerical results (solid lines) with the glory 
approximation (dashed lines)
given by Eq.~(\ref{Glory electro}), obtaining excellent 
agreement for $\theta \approx \pi$.}
 \label{electro_scs-fig}
\end{figure}

For small angles, this scattering cross section is the same for the
massless scalar and electromagnetic fields, 
and it is given by~\cite{Mashhoon1973, Fabbri1975}
\begin{equation}
 \frac{d\sigma}{d\Omega} \approx \frac{16 M^2}{\theta^4}.
 \label{wf_scs}
\end{equation}
In fact, the same behavior for small angles is obtained for massless 
fermionic and gravitational fields scattered in 
Schwarzschild spacetime~\cite{Futterman, Dolan2006, Dolan2008a}.

The glory approximation for scattering of electromagnetic waves 
by a Schwarzschild black hole can be 
determined using the strong field approximation for the deflection angle 
(which was first obtained by Darwin~\cite{Darwin}) together with the general 
glory formula~\cite{MMNZ}, namely
\begin{equation}
 \left.\frac{d\sigma}{d\Omega}\right|_{\theta\approx\pi} \approx
 2\pi\omega b_{g}^{2} \left|\frac{db}{d\theta}\right|_{\theta=\pi}
 [J_{2s}(\omega b_{g} \sin\theta)]^2,
 \label{Gen glory}
\end{equation}
where $b$ is the impact parameter of the incident particle, 
$J_l (x)$ are the Bessel functions of first kind,
$b_{g}$ is the impact parameter for which the scattering 
angle is $\pi$, and $s$ is the particle spin.
For the electromagnetic field ($s=1$), the glory scattering cross section 
is given by
\begin{equation}
 \left.\frac{1}{M^2}\frac{d\sigma}{d\Omega}\right|_{\theta\approx\pi}
 \approx 30.75 M\omega [J_{2}(5.36 M\omega \sin\theta)]^2,
 \label{Glory electro}
\end{equation}
where the coefficients ($30.75$ and $5.36$) were obtained by solving 
the geodesic equation numerically~\cite{CDO}.

In order to evaluate the electromagnetic scattering cross section 
numerically, we first solve Eq.~(\ref{Radial eq}) and 
match the solution with Eqs.~(\ref{varphi at infinity})
and~(\ref{Phase shift}).
The scattering cross section for arbitrary frequencies 
and angles is obtained through Eq.~(\ref{electro_scs}).
The numerical method used here is analogous to the one 
described in Ref.~\cite{DOC}.
We have employed an iterative method similar to that 
used in Refs.~\cite{YRW, Dolan2008} to improve the numerical 
convergence of the partial wave series.

In Fig.~\ref{electro_scs-fig} we show the differential electromagnetic 
scattering cross section of Schwarzschild black holes computed 
numerically for different values of the incident wave frequency
($M \omega = 1, 2, 3, 4$).
We also show the results for the glory scattering 
[given by Eq.~(\ref{Glory electro})] in each case.
Our numerical results are in excellent agreement with the 
glory approximation for $\theta \approx \pi$.

\begin{figure}
\includegraphics[scale=1]{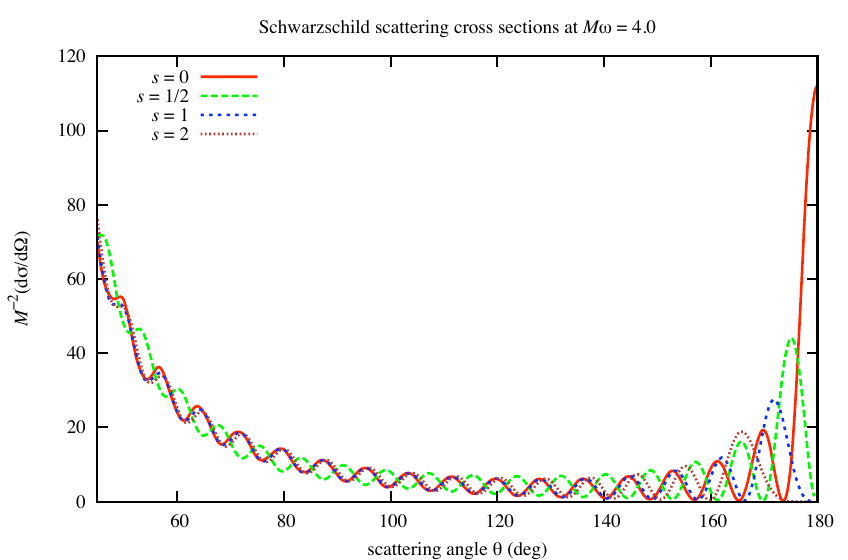}
\includegraphics[scale=1]{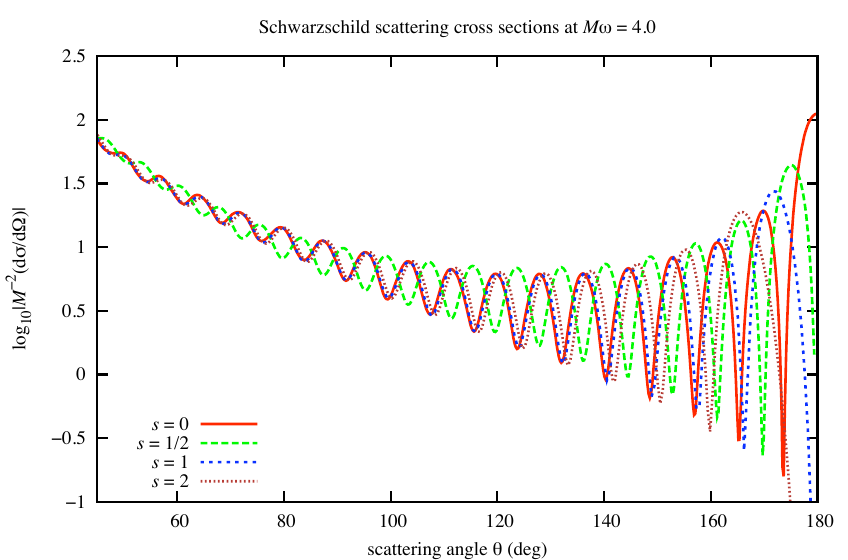}
\caption{Scattering cross section of Schwarzschild black
holes for massless scalar ($s=0$), electromagnetic ($s=1$), 
gravitational ($s=2$) and massless fermionic fields ($s=1/2$)
at $M \omega = 4.0$. 
Note the log scale on the vertical axis of the lower plot.
We see that, as all other non-zero spin fields, the 
electromagnetic wave has a vanishing scattering cross section
in the backward direction.}
 \label{schw_scs-s0s1}
\end{figure}

The zero in the backward direction (Fig.~\ref{electro_scs-fig}) is a consequence of the parallel-transport of the polarization vector along a geodesic. Consider an incoming geodesic ray in the z-direction which orbits the hole once, to return in the opposite direction ($\theta = \pi$). Assume, without loss of generality, it has an electric-field vector in the $x$-direction. If the ray orbits in the $x$-$z$ plane then the vector will be reversed, whereas if the ray orbits in the $y$-$z$ plane the vector remains unchanged. Hence, by integrating over the circular degeneracy (all orbital planes), there is perfect cancellation. Similar arguments hold for other spins \cite{Zhang, Futterman}.

In Fig.~\ref{schw_scs-s0s1} we plot the differential scattering cross
section of Schwarzschild black holes for massless scalar ($s=0$), 
massless spinor ($s=1/2$),
electromagnetic ($s=1$) and gravitational ($s=2$) waves.
As expected, in the backward direction all non-zero spin massless fields 
have vanishing cross section, whereas the zero-spin (scalar) massless field 
has a glory maximum at $\theta = \pi$.
We see that scalar ($s=0$) and electromagnetic ($s=1$) scattering 
cross sections are very similar in the angular 
region $45^\circ<\theta<160^\circ$. All integer spin fields ($s=0,1,2$) 
behave similarly for $45^\circ<\theta<120^\circ$. 
Bosonic ($s=0$, $1$, $2$) and fermionic ($s=1/2$) scattering cross sections 
oscillate in antiphase throughout almost all the angular range 
of Fig.~\ref{schw_scs-s0s1} (except near $\theta \sim \pi$).

The regular oscillations in the cross sections of Fig. \ref{schw_scs-s0s1} can be understood semi-classically. They arise from the interference of rays passing in opposite senses around the hole. There is a `path difference' between the rays passing through angles $\theta$ and $2\pi - \theta$. A maximum (minimum) occurs when the path difference is an integer (half-integer) multiple of the wavelength $\lambda$. Hence the angular width of the oscillations is inversely proportional to $M \omega$.

%
%
In summary, we have studied the scattering of a monochromatic planar electromagnetic 
wave by a Schwarzschild black hole.
We have applied the partial wave method to obtain the differential
scattering cross section numerically, for different values of the 
frequency of the incident plane wave and for different values of 
the scattering angle. We have presented graphs with accurate numerical results 
for massless fields of all spins, that is, scalar ($s=0$), fermionic ($s=1/2$), 
electromagnetic ($s=1$) and gravitational ($s=2$) fields.
All non-zero spin massless fields have a vanishing scattering 
cross section in the backward direction ($\theta=\pi$), whereas the scattering 
cross section of the massless scalar field has a local maximum.

\begin{acknowledgments}
The authors would like to thank Conselho Nacional de Desenvolvimento 
Cient\'\i fico e Tecnol\'ogico (CNPq) for partial financial support, 
and Roberto Fabbri for email correspondence. 
S. D. acknowledges financial support from the Irish Research Council 
for Science, Engineering and Technology (IRCSET).
S. D. and E. O. thank the Universidade Federal do Par\'a (UFPA) 
in Bel\'em for kind hospitality. 
L. C. and E. O. acknowledge partial financial 
support from Coordena\c{c}\~ao de Aperfei\c{c}oamento de Pessoal
de N\'\i vel Superior (CAPES).
\end{acknowledgments}

\end{document}